\documentclass[conference,letterpaper]{IEEEtran}
\IEEEoverridecommandlockouts
\usepackage[utf8]{inputenc} 
\usepackage[T1]{fontenc}
\usepackage{url}
\usepackage{ifthen}
\usepackage{cite}
\usepackage[cmex10]{amsmath} 

\usepackage{amssymb,amsfonts}
\usepackage{algorithm,algorithmic}
\usepackage{graphicx}
\usepackage{textcomp}
\usepackage{xcolor}
\usepackage{url}
\usepackage{color}
\usepackage{bm}
\usepackage{subfigure}

\DeclareMathOperator*{\argmin}{\arg\!\min}
\newtheorem{Thm}{Theorem}
\newtheorem{Lem}{Lemma}

\newtheorem{Prob}{Problem}

\newcommand{\wqr}{\textcolor{black}}
\newcommand{\wqb}{\textcolor{black}}
\newcommand{\wqc}{\textcolor{black}}
\newcommand{\wqm}{\textcolor{black}}

\newcommand{\nwqc}{\textcolor{black}}
\newcommand{\Globecomr}{\textcolor{black}}
\newcommand{\Globecomb}{\textcolor{black}}

\newcommand{\Acceptr}{\textcolor{black}}
\newcommand{\Acceptb}{\textcolor{black}}

\newcommand{\eqa}{\overset{(a)}{=}}
\newcommand{\eqb}{\overset{(b)}{=}}

\newcommand{\eqd}{\overset{(d)}{=}}

\newcommand{\llc}{\overset{(c)}{<}}

\begin{document}
\title{Optimization-based Block Coordinate Gradient Coding
} 



\author{%
  \IEEEauthorblockN{Qi Wang,\ Ying Cui,\ Chenglin Li,\ Junni Zou,\ Hongkai Xiong}
  \IEEEauthorblockA{\textit{Dept. of EE,\ Shanghai Jiao Tong University,\ China}\\
  \textit{\{wang\_qi,\ cuiying,\  lcl1985,\ zoujunni,\ xionghongkai\}@sjtu.edu.cn}}
  \thanks{\Acceptr{This work was supported in part by the National Natural Science Foundation of China under Grant 61871267, Grant 61831018, Grant 61931023, Grant 61972256, Grant 61971285, Grant 61720106001, Grant 61932022, and in part by the Program of Shanghai Science and Technology Innovation Project under Grant 20511100100.
  This work is to appear in Proc. of IEEE GLOBECOM, 2021.}}
}


\maketitle

\begin{abstract}
	Existing gradient coding schemes introduce identical redundancy across the coordinates of gradients and hence cannot fully utilize the computation results from partial stragglers. This motivates the introduction of diverse redundancies across the coordinates of gradients. This paper considers a distributed computation system consisting of one master and $N$ workers characterized by a general partial straggler model and focuses on solving a general large-scale machine learning problem with $L$ model parameters. We show that it is sufficient to provide at most $N$ levels of redundancies for tolerating $0, 1,\cdots, N-1$ stragglers, respectively. Consequently, we propose an optimal block coordinate gradient coding scheme based on a stochastic optimization problem that optimizes the partition of the $L$ coordinates into $N$ blocks, each with identical redundancy, to minimize the expected overall runtime for collaboratively computing the gradient. We obtain an optimal solution using a stochastic projected subgradient method and propose two low-complexity approximate solutions with closed-from expressions, for the stochastic optimization problem. We also show that under a shifted-exponential distribution, for any $L$, the expected overall runtimes of the two approximate solutions and the minimum overall runtime have sub-linear multiplicative gaps in $N$. To the best of our knowledge, this is the first work that optimizes the redundancies of gradient coding introduced across the coordinates of gradients.
\end{abstract}




\section{Introduction}
\label{sec:Introduction}

\begin{figure*}[t]
\begin{center}
  \subfigure[\scriptsize{Master-worker distributed computation system.}\label{subfig:system_master_worker}]
  {\resizebox{2.7cm}{!}{\includegraphics{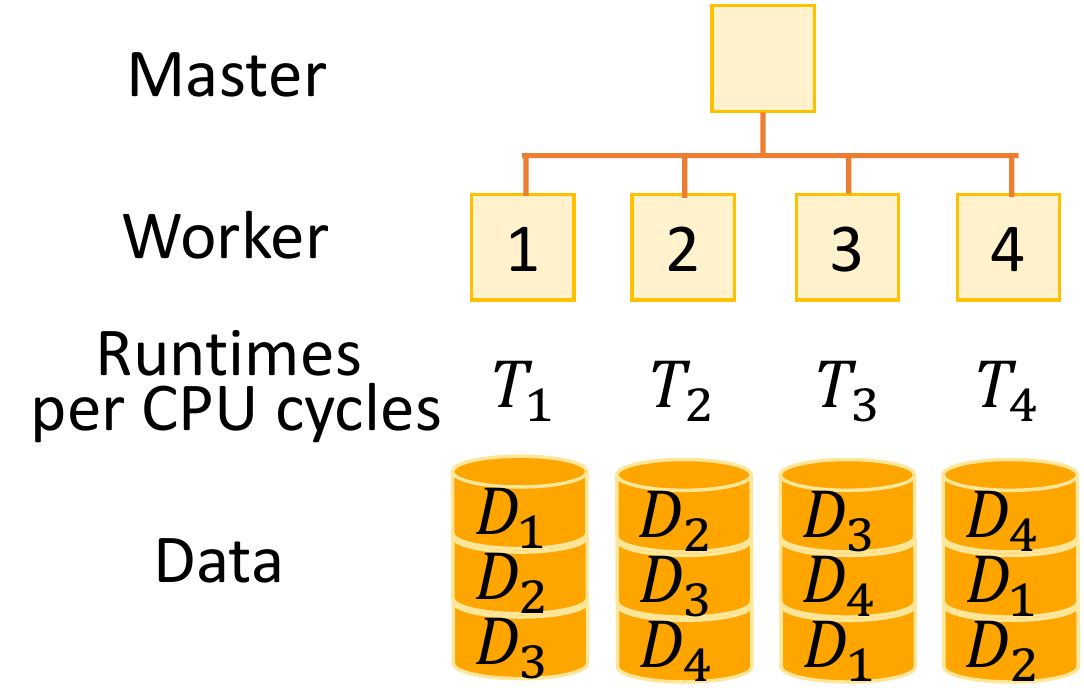}}}
  \ \
  \subfigure[\scriptsize{Gradient coding scheme in~\cite{Tandon_gradientcoding} with $s=1$. The overall runtime is $\frac{Mb}{2}T_0$.}\label{subfig:system_Tandon_1}]
  {\resizebox{4.6cm}{!}{\includegraphics{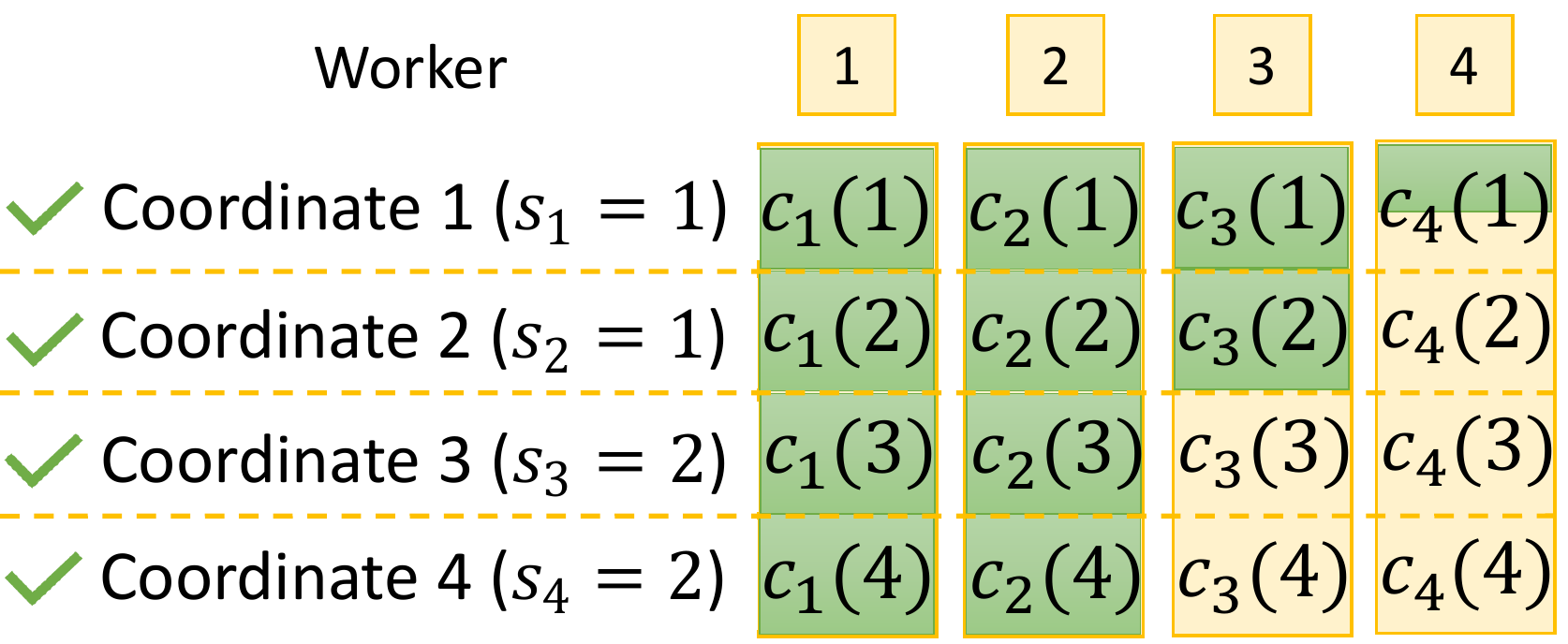}}}
  \ \
  \subfigure[\scriptsize{Gradient coding scheme in~\cite{Tandon_gradientcoding} with $s=2$. The overall runtime is $\frac{3Mb}{10}T_0$.}\label{subfig:system_Tandon_2}]
  {\resizebox{4.6cm}{!}{\includegraphics{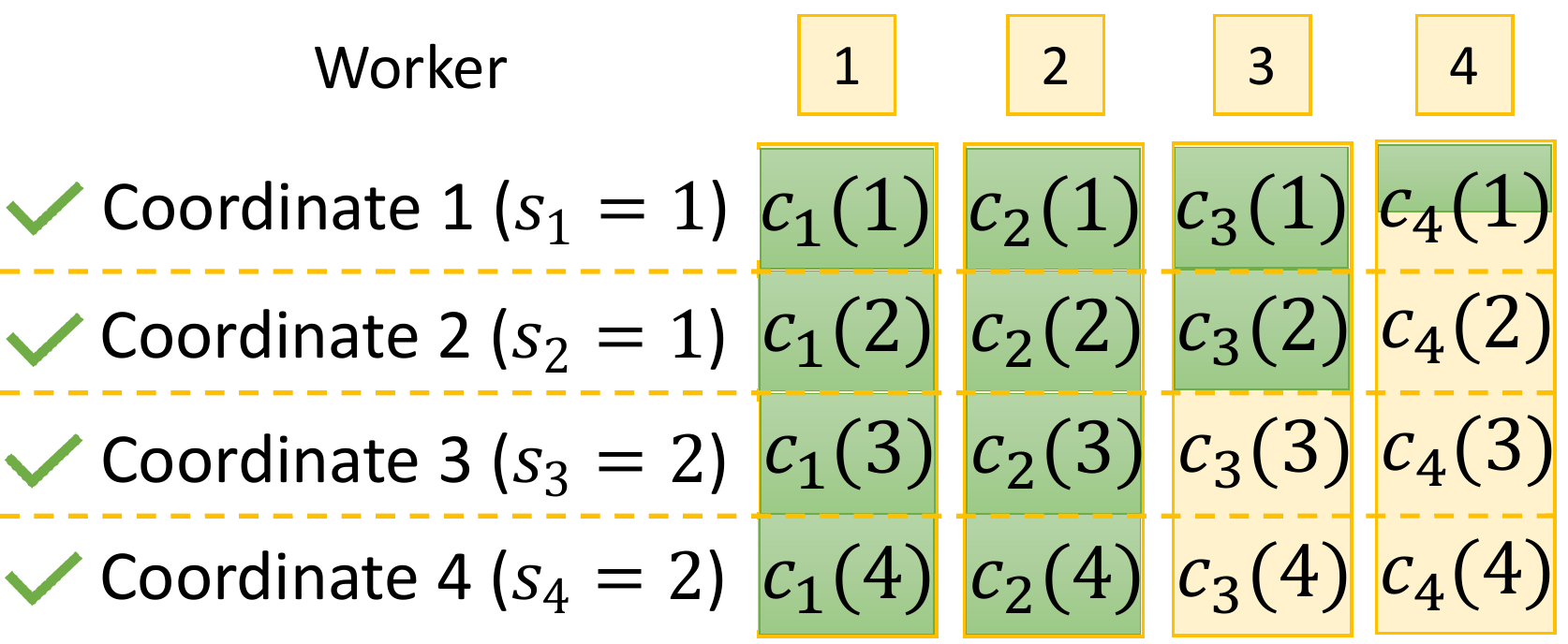}}}
  \ \
  \subfigure[\scriptsize{Proposed coordinate gradient coding scheme with $\mathbf{s}   =(1,1,2,2)$. The overall runtime is $\frac{Mb}{4}T_0$.}\label{subfig:system_proposed}]
  {\resizebox{4.6cm}{!}{\includegraphics{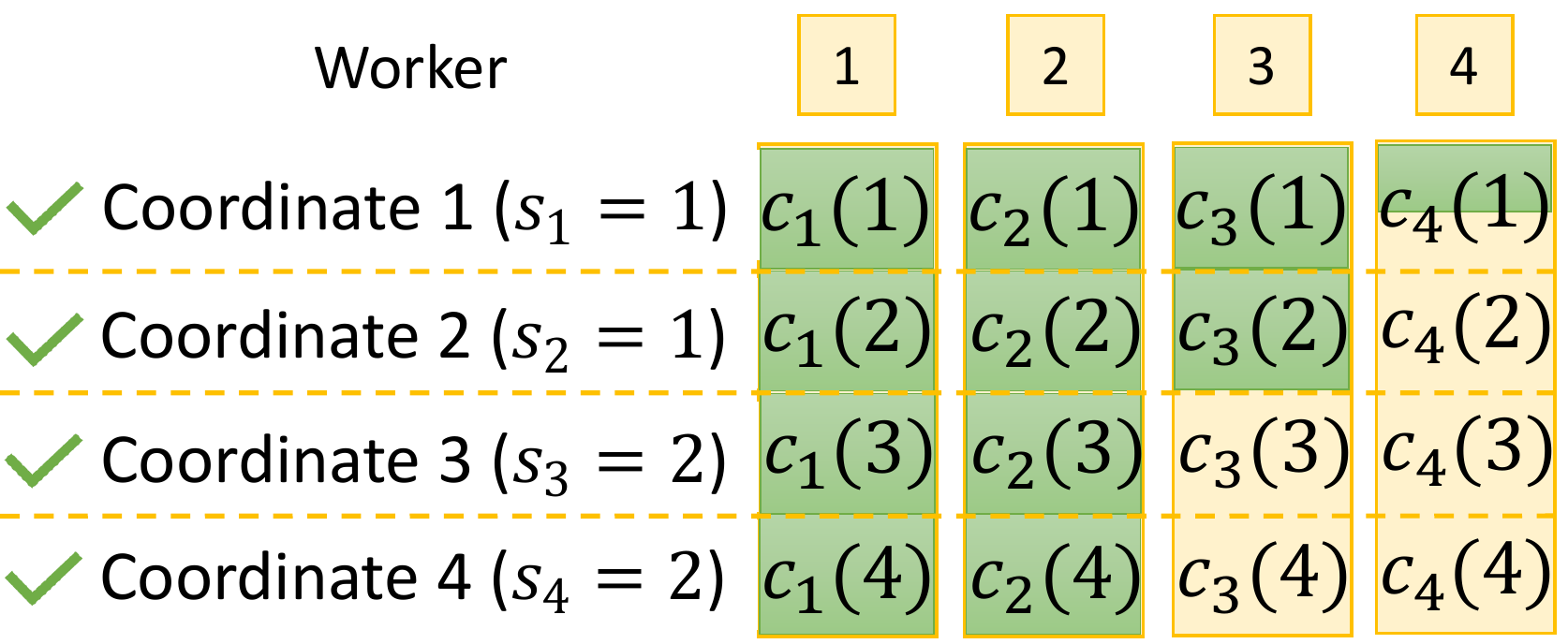}}}
  \end{center}
  \vspace{-0.5cm}
         \caption{
         Comparison between the proposed coordinate gradient coding scheme with the existing gradient coding schemes at $N=4$, $L=4$ and $\mathbf{T}=\left( \frac{1}{10},\frac{1}{10},\frac{1}{4},1 \right)T_0$.
         The green part represents the coded partial derivatives that have been computed by a worker.
         Let $c_n(l)$ denote the $l$-th coded partial derivative computed by worker $n$, for all $l = 1, ..., L$ and $n = 1, ..., N$.
         Here $c_1(l) = g_1(l)-g_2(l),\ c_2(l) = g_2(l)+g_3(l),\ c_3(l) = g_3(l) - g_4(l),\ c_4(l) = g_1(l) + g_4(l)$, for all $l\in\{l|s_l=1\}$, 
         and $c_1(l) = g_1(l) + \frac{1}{3}g_2(l) +\frac{2}{3}g_3(l),\ c_2(l) = g_2(l) + \frac{1}{2}g_3(l) + \frac{3}{2}g_4(l),\ c_3(l) = 2g_1(l) + g_3(l) - g_4(l),\ c_4(l) = -\frac{1}{2}g_1(l)+\frac{1}{2}g_2(l)+g_4(l)$, for all $l\in\{l|s_l=2\}$,
         where $g_n(l)\triangleq \frac{\partial F(\mathcal{D}_n;\bm{\theta})}{\partial\theta_l}$ for all $n\in[N]$ and $l\in[L]$\wqb{~\cite{Tandon_gradientcoding}}.
	}\label{fig:motivating_example}
	\vspace{-0.6cm}
\end{figure*}

Due to the explosion in the numbers of samples and features of modern datasets, it is impossible to train a model by solving a large-scale machine learning problem on a single node. This challenge naturally leads to distributed learning in a master-worker distributed computation system. However, slow workers, also referred to as \emph{stragglers}, can significantly affect computation efficiency. Generally speaking, there exist two straggler models. One is the full (persistent) straggler model where stragglers are unavailable permanently \cite{Tandon_gradientcoding,Tandon_approx,Bitar_approx}. The other is the partial (non-persistent) straggler model, where stragglers are slow but can conduct a certain amount of work \cite{Min_Ye_communication,   Lee_speeding_up,   Yu_polynomial,Draper_exploitation,Draper_hierarchical_ISIT,Draper_hierarchical_CSIT,Gunduz_approx_1,Gunduz_approx_2   }. The partial straggler model is more general than the full straggler model, as the former with a Bernoulli distribution for each worker's computing time degenerates to the latter.

Recently, several coding-based distributed computation techniques have been developed to mitigate the effect of stragglers in training the model via gradient descent algorithms. The common idea is to enable robust collaborative computation of a gradient in the general form \cite{Tandon_gradientcoding,Tandon_approx,Min_Ye_communication,   Bitar_approx   } or some of its components \cite{Lee_speeding_up,   Yu_polynomial,Draper_exploitation,Draper_hierarchical_ISIT,Draper_hierarchical_CSIT,    Gunduz_approx_1,Gunduz_approx_2} in the presence of stragglers. 
More specifically, \cite{Tandon_gradientcoding,Tandon_approx,Min_Ye_communication,        Bitar_approx   } propose \emph{gradient coding} schemes \Globecomb{\cite{Tandon_gradientcoding,Tandon_approx,Min_Ye_communication} or \emph{approximate gradient coding} schemes \cite{Bitar_approx}} for \Globecomb{exactly or approximately} calculating a general gradient under the full straggler model \cite{Tandon_gradientcoding,Tandon_approx,     Bitar_approx   }, $\alpha$-partial straggler model \cite{Tandon_gradientcoding}, and partial straggler model \cite{Min_Ye_communication}; \cite{Lee_speeding_up,   Yu_polynomial,Draper_exploitation,Draper_hierarchical_ISIT,Draper_hierarchical_CSIT,   Gunduz_approx_1,Gunduz_approx_2} propose \emph{coded computation} schemes for calculating matrix multiplication under the full straggler model \cite{Yu_polynomial} and partial straggler model \cite{Lee_speeding_up,   Draper_exploitation,Draper_hierarchical_ISIT,Draper_hierarchical_CSIT,Gunduz_approx_1,Gunduz_approx_2}. 
In addition, \cite{Draper_hierarchical_ISIT,Draper_hierarchical_CSIT} optimize the coding parameters under the partial straggler model with the computing times of workers independent and identically distributed (i.i.d.) according to a shifted-exponential distribution.
\Globecomb{Note that gradient coding and coded computation are used for gradient descent (GD) methods, while approximate gradient coding is used for stochastic gradient descent (SGD) methods.
Furthermore, note that SGD has a weaker convergence guarantee than GD.
To achieve the strongest convergence guarantee, we focus on gradient coding and coded computation which apply to GD.}


Towards a broader range of applications and a stronger convergence  guarantee, this paper focuses on designing exact gradient coding schemes under a general partial straggler model with the computing times of workers following an arbitrary distribution in an i.i.d. manner. 
The gradient coding schemes in \cite{Tandon_gradientcoding,Tandon_approx,Min_Ye_communication} introduce identical redundancy across the coordinates of gradients (for a partition of the whole data set).
Hence, these schemes cannot fully utilize the coordinates of coded gradients computed by stragglers under the partial straggler model.
This motivates us to optimally diversify the redundancies across the coordinates of gradients by designing coding parameters to effectively utilize the computational resource of all workers. Note that the optimization of the coding parameters for coded matrix multiplication in \cite{Draper_hierarchical_ISIT,Draper_hierarchical_CSIT} does not apply to gradient coding.

This paper considers a distributed computation system consisting of one master and $N$ workers characterized by a general partial straggler model and focuses on solving a general large-scale machine learning problem with $L$ model parameters. First, we propose a coordinate gradient coding scheme with $L$ coding parameters, each controlling the redundancy for one coordinate, to maximally diversify the redundancies across the coordinates of gradients. Then, we formulate an optimization problem to minimize the expected overall runtime for collaboratively computing the gradient by optimizing the $L$ coding parameters for coordinates. The problem is a challenging stochastic optimization problem with a large number ($L$) of variables. Next, we convert the original problem with $L$ coding parameters for the $L$ coordinates to an equivalent but a much simpler problem with $N$ coding parameters for $N$ blocks of coordinates. This equivalence implies that it is sufficient to provide at most $N$ levels of redundancies for tolerating $0, 1,\cdots, N-1$ stragglers, respectively, and it remains to optimally partition the $L$ coordinates into $N$ blocks, each with identical redundancy. 
We obtain an optimal solution of the simplified problem using a stochastic projected subgradient method and propose two low-complexity approximate solutions with closed-from expressions. We also show that under a shifted-exponential distribution, \wqm{for any $L$}, the expected overall runtimes of the two approximate solutions and the minimum overall runtime have sub-linear multiplicative gaps \wqm{in $N$}. Finally, numerical results show that the proposed solutions significantly outperform existing ones, and the proposed approximate solutions achieve close-to-minimum expected overall runtimes.

\vspace{-0.4cm}
\section{System Setting}
\label{sec:System_Setting}
\vspace{-0.2cm} 

As illustrated in Fig.~\ref{subfig:system_master_worker}, we consider a master-worker distributed computation system consisting of one master and $N$ workers~\cite{Tandon_gradientcoding}.
Let $[N]\triangleq\{1,\cdots,N\}$ denote the set of worker indices.
The master and workers have computation and communication capabilities.
We assume that the master and each worker are connected by a fast communication link and hence we omit the communication time as in \cite{Draper_exploitation,Draper_hierarchical_ISIT}.
We consider a general partial straggler model for the workers:
at any instant, the CPU cycle times of the $N$ workers, denoted by $T_n,n\in [N]$, are i.i.d. random variables;
the values of $T_n,n\in [N]$ at each instant are not known to the master but the distribution is known to the master.
Notice that the adopted straggler model includes those in~\cite{Lee_speeding_up,   Draper_hierarchical_ISIT,Draper_hierarchical_CSIT,Min_Ye_communication} as special cases.
Besides, note that most theoretical results in this paper do not require any assumption on the distribution of $T_n, n\in[N]$. 
Let $T_{(1)},T_{(2)},\cdots,T_{(N)}$ be $T_{n},n\in[N]$ arranged in increasing order, so that $T_{(n)}$ is the $n$-th smallest one.
As in \cite{Tandon_gradientcoding,Min_Ye_communication,Tandon_approx   }, we focus on the following distributed computation scenario. 
The master holds a data set of $M$ samples, denoted by $\mathcal{D}$, and aims to train a model with model parameters $\bm{\theta}\in\mathbb{R}^L$ by solving the following machine learning problem:
\begin{equation}
\setlength{\abovedisplayskip}{3pt}
\setlength{\belowdisplayskip}{1pt}
	\min_{\mathbf{\bm{\theta}}} F(\mathcal{D};\mathbf{\bm{\theta}}) \triangleq \sum_{\mathbf{y}\in\mathcal{D}} f(\mathbf{y};\bm{\theta})\nonumber
\end{equation}
using commonly used gradient descent methods,\footnote{The proposed scheme can also apply to stochastic gradient descent methods as discussed in~\cite{Min_Ye_communication}.} with the $N$ workers and master collaboratively computing the gradient $\nabla_{\bm{\theta}} F(\mathcal{D};\bm{\theta})=\sum_{\mathbf{y}\in\mathcal{D}}\nabla_{\bm{\theta}}f(\mathbf{y};\bm{\theta})$ in each iteration of the gradient descent.
Notice that the model size $L$ is usually much larger than the number of workers $N$.
In this paper, we consider a general differentiable function $f(\cdot)$.


\Acceptb{We focus on the case where each worker sequentially computes the coordinates and sends each coordinate to the master once its computation is completed.\footnote{The considered system model can be readily extended to the case that the machine learning problem is solved using a neural network by changing the basic unit from one coordinate to a block of coordinates which associate with one layer of the neural network.}
In this case, 
the gradient coding schemes proposed in \cite{Tandon_gradientcoding,Min_Ye_communication,Tandon_approx   }, which introduce redundancy across the coordinates of gradients (for a partition of the whole data set) cannot efficiently mitigate the impact of full or partial stragglers.
This is because they cannot utilize the computation results from partial stragglers as illustrated in Fig.~\ref{fig:motivating_example}.
In particular, in Fig.~\ref{subfig:system_Tandon_1}, the computation results of coordinates 3, 4 from worker 1, coordinates 3, 4 from worker 2, and  coordinate 1 from worker 4 are not utilized;
in Fig.~\ref{subfig:system_Tandon_2}, the coded partial derivative 4 from worker 1, coded partial derivative 4 from worker 2, coded partial derivatives 1, 2 from worker 3, and coded partial derivative 1 from worker 4 are not utilized.
This motivates us to introduce diverse redundancies across the coordinates of gradient to effectively utilize the computation resource.}

\section{Coordinate Gradient Coding}
\label{sec:Coordinate_Gradient_Coding}

To diversify the redundancies across the coordinates of gradients (for a partition of the whole data set), i.e.,  partial derivatives, we propose a coordinate gradient coding scheme parameterized by $\mathbf{s}\triangleq (s_l)_{l\in [L]}$, 
where the coding parameters $\mathbf{s}$ for the $L$ coordinates satisfy:\footnote{For ease of exposition, we present the gradient coding scheme for the case that the machine learning problem is directly solved by GD.
The proposed gradient coding scheme can be readily extended to the case that the machine learning problem is solved using a neural network by changing the basic unit from one coordinate to a block of coordinates which associate with one layer of the neural network.}
\begin{equation}
\setlength{\abovedisplayskip}{2pt}
\setlength{\belowdisplayskip}{2pt}
	s_l \in \{0,\cdots,N-1\},\ l\in [L].\label{equ:constraint_s_1}
\end{equation}
That is, the master can tolerate $s_l$ stragglers in recovering the $l$-th partial derivative $\frac{\partial F(\mathcal{D};\bm{\theta})}{\partial \theta_l}$.
Note that $s_l=0$ means that no redundancy is introduced across the $l$-th partial derivatives.
The proposed coordinate gradient coding scheme generalizes the gradient coding scheme in~\cite{Tandon_gradientcoding}, where $s_l,l\in[L]$ are identical.
\Acceptb{Later, we investigate the optimization of the coding parameter $\mathbf{s}$ for the $L$ coordinates under the constraints in~\eqref{equ:constraint_s_1}, from which we can see that the optimization-based coordinate gradient coding schemes become block coordinate gradient coding schemes which can be easily implemented in practice.
The proposed scheme operates in two phases.}

\textbf{Sample Allocation Phase}:
First, the master partitions dataset $\mathcal{D}$ into $N$ subsets of size $\frac{M}{N}$, denoted by $\mathcal{D}_i,i\in\ [N]$~\cite{Tandon_gradientcoding,Min_Ye_communication,Tandon_approx   }.
Then, for all $n\in [N]$, the master allocates the $\max_{l\in [L]}s_l+1$ subsets, $\mathcal{D}_i,i\in \mathcal{I}_n\triangleq \{j\oplus(n-1)|j\in[\max_{l\in [L]}s_l+1] \}$, to worker $n$, where the operator $\oplus$ over set $[N]$ is defined as: {\small{$a_1 \oplus a_2 \triangleq \left\{
	\begin{aligned}
		a_1+a_2,\quad {\rm if}\ a_1+a_2 \le N\\
		a_1+a_2-N,\quad {\rm if}\ a_1+a_2 > N
	\end{aligned}
	\right.$, for all $a_1,a_2\in[N]$.}}
Note that the master is not aware of the values of $T_n,n\in[N]$ in the sample allocation phase.
	
\textbf{Collaborative Training Phase}:
In each iteration, the master first sends the latest $\bm{\theta}$ to all workers.
Then, for $l\in[L]$, \Acceptb{subsequent procedures are conducted.}
Each worker $n\in[N]$ computes the \wqc{$l$-th coded partial derivative} based on the encoding matrix in~\cite{Tandon_gradientcoding} (with $s=s_l$)
and sends it to the master. 
The master sequentially receives the \wqc{coded partial derivatives} from each worker and recovers $\frac{\partial F(\mathcal{D};\bm{\theta})}{\partial \theta_l}$, based on the decoding matrix in~\cite{Tandon_gradientcoding} (with $s=s_l$) once it receives the \wqc{$l$-th coded partial derivatives} from the $N-s_l$ fastest workers with CPU cycle times $T_{(n)}, n\in[N-s_l]$. 
Note that the orders for computing and sending the coded partial derivatives are both $1,\cdots,L$.
Once the master has recovered $\frac{\partial F(\mathcal{D};\bm{\theta})}{\partial \theta_l}$ for all $l\in[L]$, it can obtain the gradient $\nabla_{\bm{\theta}} F(\mathcal{D};\bm{\theta})$~\cite{Tandon_gradientcoding}.  

\Globecomb{Let $b$ denote the maximum of the numbers of CPU cycles for computing $\frac{\partial F(\mathcal{D};\bm{\theta})}{\partial \theta_l}$, $l\in[L]$.}
\Globecomb{For tractability, in this paper we use the maximum, $b$, in optimizing the coding parameters.}\footnote{\Globecomb{The proposed optimization framework can be extended to consider \Acceptb{exact} numbers of CPU cycles for computing $\frac{\partial F(\mathcal{D};\bm{\theta})}{\partial \theta_l},l\in[L]$, \Acceptb{in optimizing the coding parameters}.}}
We omit the computation loads for encoding at each worker and decoding at the master as they are usually much smaller than the computation load for calculating the partial derivatives at a large number of samples in practice.
Thus, for all $n\in[N]$ and $l\in[L]$, the completion time for computing the \wqc{$l$-th coded partial derivative} at worker $n$ is $\frac{M}{N}bT_n\sum_{i=1}^l\left(s_i+1\right)$.
For all $l\in[L]$, the completion time for recovering $\frac{\partial F(\mathcal{D};\bm{\theta})}{\partial \theta_l}$ at the master is $\frac{M}{N}bT_{(N-s_l)}\sum_{i=1}^l\left(s_i+1\right)$.
Therefore, the overall runtime for the workers and master to collaboratively compute the gradient $\nabla_{\bm{\theta}} F(\mathcal{D};\bm{\theta})$ is:
\begin{equation}
\setlength{\abovedisplayskip}{2pt}
\setlength{\belowdisplayskip}{2pt}
	\tau(\mathbf{s,T}) = \frac{M}{N}b \max_{l\in [L]}\left\{T_{(N-s_l)} \sum_{i=1}^l\left(s_i+1\right)\right\},\label{equ:tau_calculate}	
\end{equation}
where $\mathbf{T}\triangleq \left(T_n\right)_{n\in [N]}$.
Note that $\tau(\mathbf{s,T})$ is a function of the parameters $\mathbf{s}$ and random vector $\mathbf{T}$ and hence is also random.
In Fig.~\ref{subfig:system_proposed}, we provide an example to illustrate the idea of the proposed coordinate gradient coding scheme.
\Acceptb{Specifically, from the example in Fig.~\ref{subfig:system_proposed}}, we can see that the proposed coordinate gradient coding scheme with coding parameters $\mathbf{s}=(1,1,2,2)$ has a shorter overall runtime than the gradient coding scheme in~\cite{Tandon_gradientcoding} with coding parameter $s=1$ or $s=2$ at $\mathbf{T}=\left( \frac{1}{10},\frac{1}{10},\frac{1}{4},1 \right)$, \Acceptb{as more computation results from partial stragglers are utilized}.

\section{Optimization-based Block Coordinate Gradient Coding}
\label{sec:Optimization_of_Coding_Parameters}

The expected overall runtime $\mathbb{E}_{\mathbf{T}}\left[\tau(\mathbf{s,T})\right]$ measures on average how fast the gradient descent-based model training can be completed in the distributed computation system with the proposed coordinate gradient coding scheme.
\hspace{-0.1cm} We would like to minimize $\mathbb{E}_{\mathbf{T}}\left[\tau(\mathbf{s,T})\right]$ by optimizing the coding parameters $\mathbf{s}$ for the $L$ coordinates under the constraints in~\eqref{equ:constraint_s_1}.
\begin{Prob}[Coding Parameter Optimization]\label{prob:original_prob}
	\begin{align}
		\tau_{\rm avg}^* \triangleq \min_{\mathbf{s}} \ \  &\mathbb{E}_\mathbf{T}\left[ \tau(\mathbf{s,T}) \right] \nonumber\\
		  \rm{s.t.}\ \  &\eqref{equ:constraint_s_1}.\nonumber
	\end{align}
\end{Prob}
\vspace{-0.1cm}

In general, the objective function does not have an analytical expression, and the model size $L$ is usually quite large.
Thus, Problem~\ref{prob:original_prob} is a challenging stochastic optimization problem.
First, we characterize the monotonicity of an optimal solution of Problem~\ref{prob:original_prob}.
\begin{Lem}[Monotonicity of Optimal Solution of Problem~\ref{prob:original_prob}]\label{lem:monotonicity}
	An optimal solution $\mathbf{s^*}\triangleq (s^*_l)_{l\in [L]}$ of Problem~\ref{prob:original_prob} satisfies $ s^*_1 \le s^*_2 \le \cdots \le s^*_L $.
\end{Lem}

\begin{IEEEproof}
	We prove Lemma~\ref{lem:monotonicity} by contradiction.
	Suppose that for any optimal solution $\mathbf{s^*}$, there exists $k\in [L]$ such that $s^*_k > s^*_{k+1}$.
	Construct a feasible solution $\tilde{\mathbf{s}} \triangleq (\tilde{s}_l)_{l\in [L]}$, where $\tilde{s}_l = s^*_l, l\neq k$ and $\tilde{s}_l = s^*_{k+1}, l=k$.
	By~\eqref{equ:tau_calculate} and the definition of $\tilde{\mathbf{s}}$, we have $\tau(\tilde{\mathbf{s}},\mathbf{T}) < \tau(\mathbf{s^*,T})$, for all $l\in [m-1]$, \Acceptb{$\{m+1,m+2,\cdots,N\}$} and $l=m$.
	This indicates that $\mathbf{s}^*$ is not an optimal solution, which contradicts with the assumption.
\end{IEEEproof}

\begin{figure}[h]
\begin{center}
  {\resizebox{3cm}{!}{\includegraphics{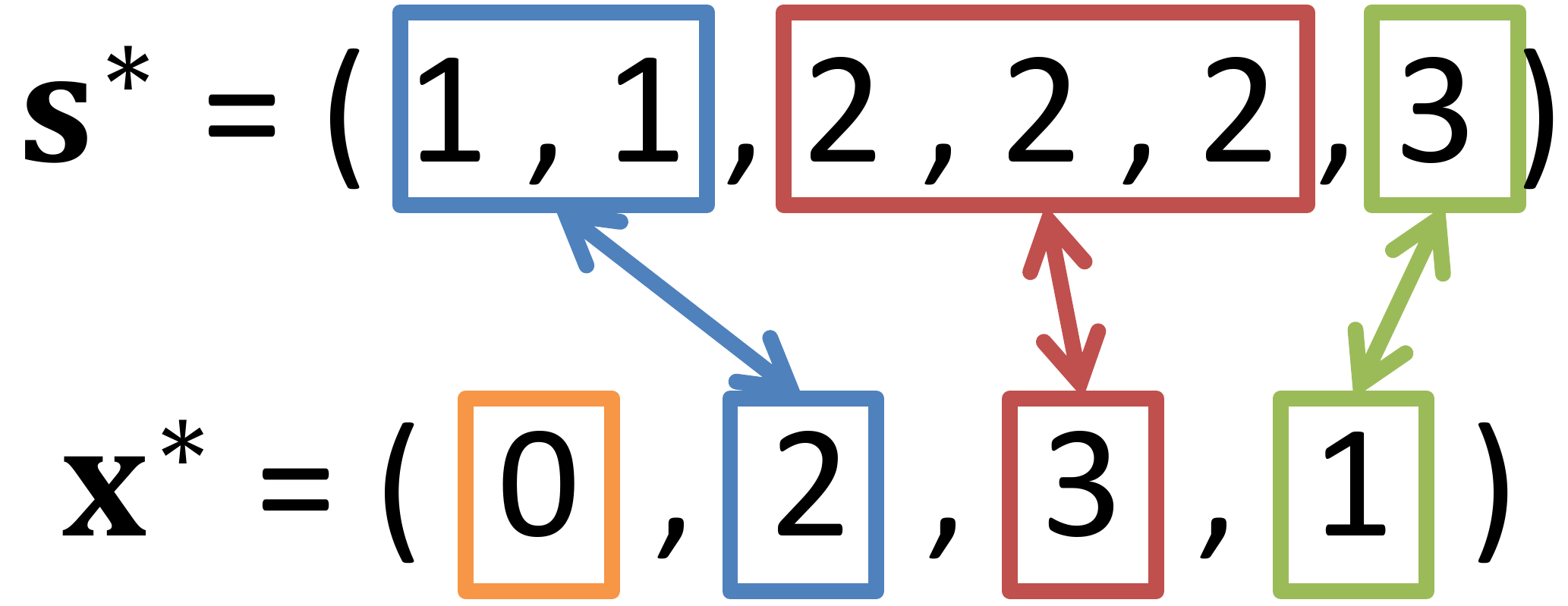}}}
  \quad\quad 
  {\resizebox{3cm}{!}{\includegraphics{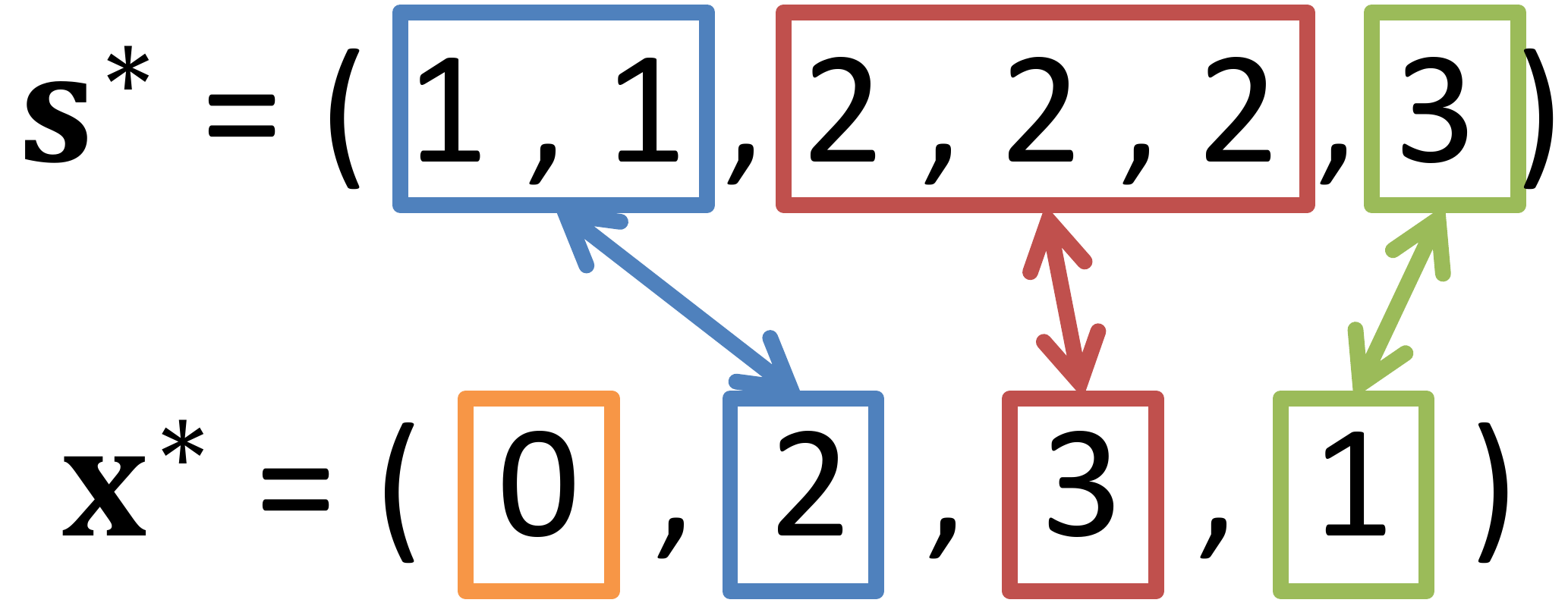}}}
  \end{center}
  \vspace{-0.3cm}
         \caption{Illustration of relation between $\mathbf{s}^*$ and $\mathbf{x}^*$ at $N=4$ and $L=6$.}\label{fig:example_illustration}
         \vspace{-0.2cm}
\end{figure} 

	Lemma~\ref{lem:monotonicity} indicates that it is sufficient to provide at most $N$ levels of redundancies for tolerating $0, 1,\cdots, N-1$ stragglers, respectively.
	\Acceptb{Thus,} it remains to optimally partition the $L$ coordinates into $N$ blocks, each with identical redundancy.
	That is, the proposed coordinate gradient coding scheme with $\mathbf{s}^*$ becomes a block coordinate gradient coding scheme.

Next, based on Lemma~\ref{lem:monotonicity}, we transform Problem~\ref{prob:original_prob} to an equivalent problem with $N$ variables.
\begin{Prob}[Equivalent Problem of Problem~\ref{prob:original_prob}]\label{prob:equivalent_1}
		\begin{align}
			\hat{\tau}_{\rm avg}^* \triangleq \min_{\mathbf{ x}} \ \  &\mathbb{E}_\mathbf{T}\left[ \hat{\tau}(\mathbf{ x},\mathbf{T}) \right] \nonumber\\
		  \rm{s.t.}\ \  &\sum_{n=0}^{N-1}  x_n=L,\label{equ:constraint_r_1}\\
		  				& x_n \in \mathbb{N},\quad n=0,1,\cdots,N-1,\label{equ:constraint_r_2}
		\end{align}
		where $\mathbf{ x} \hspace{-0cm} \triangleq \hspace{-0cm} ( x_n)_{n=0,\cdots,N-1}$\hspace{-0cm}, $\mathbb{N}$ denotes the set of  natural numbers, and 
		\begin{align}
			\hat{\tau}(\mathbf{ x},\mathbf{T})\ \triangleq \frac{M}{N}b\underset{n=0,\cdots,N-1}{\max}\left\{ T_{(N-n)}\sum_{i=0}^n(i+1) x_i \right\}.
		\end{align}

\end{Prob}
\begin{Thm}[Equivalence between Problem~\ref{prob:original_prob} and Problem~\ref{prob:equivalent_1}]\label{thm:prob_equvalence}
	An optimal solution of Problem~\ref{prob:original_prob}, denoted by $\mathbf{s}^*=(s^*_l)_{l\in [L]}$, and an optimal solution of Problem~\ref{prob:equivalent_1}, denoted by $\mathbf{ x}^*=( x^*_n)_{n=0,\cdots,N-1}$, satisfy:
	\begin{align}
		x_n^* &= \sum_{l\in[L]} I(s_l^*=n),\ n=0,1,\cdots,N-1,\label{equ:change_of_var_x}\\
		s_l^* &= \min\left\{ i\Big|\sum_{n=0}^{i} x_n^* \ge l \right\},\ l\in[L],\label{equ:change_of_var_s}
	\end{align}
	where $I(\cdot)$ denotes the indicator function.
	Furthermore, their optimal values, $\tau_{\rm avg}^*$ and $\hat{\tau}_{\rm avg}^*$, satisfy $\tau_{\rm avg}^*=\hat{\tau}_{\rm avg}^*$.
\end{Thm}
\begin{IEEEproof}
	Consider $\mathbf{s}^*$ satisfying $ s^*_1 \le s^*_2 \le \cdots \le s^*_L $.
	First, it is clear that the change of variables given by $x_n = \sum_{l\in[L]} I(s_l=n),\ n=0,1,\cdots,N-1$ and $s_l = \min\left\{ i\big|\sum_{n=0}^{i} x_n \ge l \right\},\ l\in[L]$ is one to one.
	Define $f(l,\mathbf{x}) \triangleq \min\left\{ i|\sum_{n=0}^i x_n \ge l \right\}$.
	\Globecomb{We treat $\sum_{i=1}^{-1} x_i\triangleq 0$ for ease of illustration.}
	Then, we have: 
	\begin{small}
	\begin{align}
\setlength{\abovedisplayskip}{0pt}
\setlength{\belowdisplayskip}{0pt}
		& \tau(\mathbf{s},\mathbf{T})
		\overset{(a)}{=}\frac{M}{N}b 
		\max_{l\in [L]}\Bigg\{T_{\left(N- f(l,\mathbf{x}) \right)}
		\sum_{i=1}^l\Bigg( f(i,\mathbf{x}) \hspace{-1mm} +1  \Bigg) \hspace{-1mm} \Bigg\} \hspace{-1mm} \nonumber\\
		&= \hspace{-0mm} \frac{M}{N}b \max_{l\in [L]}\Bigg\{ T_{\left(N- f(l,\mathbf{x}) \right)} \Bigg( \hspace{-0mm} \sum_{g=0}^{f(l,\mathbf{x})-1} \hspace{-0mm} \sum_{i=\sum_{j=0}^{g-1} x_j+1}^{\sum_{j=0}^g x_j} \hspace{-0mm} 
		\Bigg( \hspace{-1mm} f(i,\mathbf{x}) \hspace{-0.5mm} + \hspace{-0.5mm} 1 \hspace{-1mm} \Bigg) \nonumber\\
		&\quad + \hspace{-0mm} \sum_{i=\sum_{j=0}^{f(l,\mathbf{x})-1} \hspace{-0mm} x_j+1}^l \hspace{-0mm} \Bigg( \hspace{-0mm} f(i,\mathbf{x}) \hspace{-0mm} + \hspace{-0mm} 1 \hspace{-0mm} \Bigg) \hspace{-0mm} \Bigg) \hspace{-0mm} \Bigg\}\nonumber\\
		& = \hspace{-0.5mm} \frac{M}{N}b \max_{l\in [L]} \hspace{-0.5mm} \Bigg\{ \hspace{-0.5mm} T_{\left(N- f(l,\mathbf{x}) \right)} \hspace{-1mm} \Bigg( \hspace{-1mm} \sum_{g=0}^{f(l,\mathbf{x})-1} \hspace{-2mm} (g+1)x_g \hspace{-0.5mm} + \hspace{-0.5mm} (f(l,\mathbf{x}) \hspace{-0.5mm} + \hspace{-0.5mm} 1) (l- \hspace{-4mm} \sum_{j=0}^{f(l,\mathbf{x})-1} \hspace{-3mm} x_j) \hspace{-1mm} \Bigg) \hspace{-1mm} \Bigg\} \nonumber\\
		& =\frac{M}{N}b \max_{h\in\{0,\cdots,N-1\}} \max_{l\in\left\{ \sum_{i=0}^{h-1}x_i+1,\cdots,\sum_{i=0}^h x_i \right\}} \Bigg\{ T_{\left(N- f(l,\mathbf{x}) \right)}  \nonumber\\
		&\quad \times \Bigg( \sum_{g=0}^{f(l,\mathbf{x})-1} (g+1)x_g + (f(l,\mathbf{x})-1+2)(l-\sum_{j=0}^{f(l,\mathbf{x})-1}x_j) \hspace{-0.5mm} \Bigg) \hspace{-0.5mm} \Bigg\} \nonumber\\ 
		& =\frac{M}{N}b \max_{h\in\{0,\cdots,N-1\}} \Bigg\{T_{(N-h)} \sum_{g=0}^h (g+1)x_g \Bigg\}
		=\hat{\tau}(\mathbf{x,T}),\nonumber
	\end{align}
	\end{small}where $(a)$ is due to the change of variables.
	\Globecomr{Therefore, we complete the proof.}
\end{IEEEproof}

\begin{figure*}[ht]
\normalsize{
	\begin{align}
		t_{n}'\hspace{-0.0cm} 
		=\hspace{-0.0cm}  -1 \Bigg/ \Bigg(\mu (N+1-n)\binom{N}{n-1}\hspace{-0.0cm} \sum_{i=0}^{n-1}(-1)^i\binom{n-1}{i}e^{\mu t_0 (N-n+i+1)} E_i(-\mu t_0 (N-n+i+1)) \Bigg).\label{equ:t'}
	\end{align}
}\hrulefill
\end{figure*}

Fig.~\ref{fig:example_illustration} illustrates the relationship between $\mathbf{x}^*$ and $\mathbf{s}^*$ given by \eqref{equ:change_of_var_x} and \eqref{equ:change_of_var_s}.
Theorem~\ref{thm:prob_equvalence} indicates that $x^*_n$ represents the number of coordinates with identical redundancy for tolerating $n$ stragglers, and $\mathbf{x}^*$ represents the optimal partition of the $L$ coordinates into $N$ blocks, each with identical redundancy.
Thus, $\mathbf{x}^*$ specifies the optimal block coordinate gradient coding scheme.
As the number of model parameters $L$ is usually much larger than the number of workers $N$, the computational complexity can be greatly reduced if we solve Problem~\ref{prob:equivalent_1} rather than Problem~\ref{prob:original_prob}.
By relaxing the integer constraints in~\eqref{equ:constraint_r_2}, we have the following continuous relaxation of  Problem~\ref{prob:equivalent_1}.

\begin{Prob}[Relaxed Continuous Problem of Problem~\ref{prob:equivalent_1}]\label{prob:relaxed_prob}
	\begin{align}
		\hat{\tau}^*_{\rm avg-ct} \triangleq \min_{\mathbf{ x}} \ \  &\mathbb{E}_\mathbf{T}\left[ \hat{\tau}(\mathbf{ x},\mathbf{T}) \right] \nonumber\\
		  \rm{s.t.}\ \  &\eqref{equ:constraint_r_1}, \nonumber\\
		  				& x_n \ge 0,\quad n=0,1,\cdots,N-1. \label{equ:constraint_r_3}
	\end{align}
\end{Prob}

One can apply the rounding method in \cite[pp. 386]{Boyd_cvxbook} to round an optimal solution of Problem~\ref{prob:relaxed_prob} to an integer-valued feasible point of Problem~\ref{prob:equivalent_1}, which is a good approximate solution when $N\ll L$ (which is satisfied in most machine learning problems).
In the following section, we focus on solving the relaxed problem in Problem~\ref{prob:relaxed_prob}.

\section{Solutions}
\label{sec:Solution}
\vspace{-0.1cm}

\subsection{Optimal Solution}
\label{sssec:solution}

Problem~\ref{prob:relaxed_prob} is a stochastic convex problem whose objective function is the expected value of a (non-differentiable) piecewise-linear function.
An optimal solution of Problem~\ref{prob:relaxed_prob}, denoted by $\mathbf{x}^{\dag}$, can be obtained by the stochastic projected subgradient method~\cite{Boyd_stochastic_subgradient_methods}.
The main idea is to compute a noisy unbiased subgradient of the objective function and carry out a projected subgradient update based on it, at each iteration.
It can be easily verified that the projection problem has a semi-closed form solution which can be obtained by the bisection method, 
and the overall computational complexity of the stochastic projected subgradient method is $O(N^2)$.

\vspace{-0.1cm}
\subsection{Approximate Solutions}
\label{ssec:approx_solution}
\vspace{-0.1cm}

In this part, we obtain two closed-form approximate solutions which are more computationally efficient than the stochastic projected subgradient method.
First, we approximate the objective function of Problem~\ref{prob:relaxed_prob} by replacing the random vector $\mathbf{T}$ with the deterministic vector $\mathbf{t}\triangleq (t_n)_{n\in[N]}$, \wqm{where} $t_n \triangleq \mathbb{E}\left[T_{(n)}\right]$ (which can be numerically computed for a general distribution of $T_n,n\in[N]$).
\begin{Prob}[Approximation of Problem~\ref{prob:relaxed_prob} at $\mathbf{t}$]\label{prob:approx_solution_time}
	\begin{align}
		\mathbf{x}^{\rm (t)} \triangleq \argmin_{\mathbf{x}} \ \  &\hat{\tau}(\mathbf{x},\mathbf{t}) \nonumber\\
		  \rm{s.t.}\ \  &\eqref{equ:constraint_r_1},\eqref{equ:constraint_r_3}.\nonumber
	\end{align}
\end{Prob}
\begin{Thm} [Closed-form Optimal Solution of Problem~\ref{prob:approx_solution_time}]\label{thm:approx_solution_time}
\begin{small} 
	\begin{align}
		&x^{\rm (t)}_0 \hspace{-0.05cm} = \hspace{-0.05cm} \frac{1} {t_N}m^{\rm (t)} \hspace{-0.1cm},\ x^{\rm (t)}_n \hspace{-0.05cm} = \hspace{-0.05cm} \frac{1}{n+1}\hspace{-0cm}\Bigg(\hspace{-0cm}\frac{1} {t_{N-n}} \hspace{-0.05cm} - \hspace{-0cm} \frac{1} {t_{N+1-n}}\hspace{-0cm}\Bigg)\hspace{-0.05cm}m^{\rm (t)} \hspace{-0.1cm},\ n\in[N-1],
	\end{align}
\end{small}where $m^{\rm (t)} \triangleq \frac{L} {\sum_{n=1}^{N-1} \frac{1}{n(n+1)t_{N+1-n}} + \frac{1}{N t_{1}}} $.
\end{Thm}
\begin{IEEEproof}
	\Globecomb{First, we prove  $\hat{\tau}(\mathbf{x},\mathbf{t}) \ge \frac{M}{N}b m^{\rm (t)}$ for all $\mathbf{x}$ satisfying \eqref{equ:constraint_r_1} and \eqref{equ:constraint_r_3} by contradiction.
	Recall $\hat{\tau}(\mathbf{x},\mathbf{t}) = \frac{M}{N}b\underset{n=0,\cdots,N-1}{\max}\left\{ t_{N-n}\sum_{i=0}^n(i+1) x_i \right\}$.
	Suppose $\hat{\tau}(\mathbf{x},\mathbf{t}) < \frac{M}{N}b m^{\rm (t)}$.
	Define $\beta_i\triangleq \frac{1}{i(i+1)t_{N+1-i}},\ i\in[N]$.
	We have:
\begin{small}
\begin{align}
	&L  \eqa \sum_{i=0}^{N-1}x_i 
		\eqb \sum_{i=1}^N \beta_i\bigg( t_{N+1-i}\sum_{j=0}^{i-1}(j+1)x_j \bigg) 
		\Globecomr{\llc} m^{\rm (t)}\sum_{i=1}^N\beta_i \nonumber\\
	&\hspace{-1mm} \eqd \frac{L} {\sum_{i=1}^{N-1}\frac{1}{i(i+1)t_{N+1-i}} + \frac{1}{Nt_{1}}} 
		\Big( \sum_{i=1}^{N-1}\frac{1}{i(i+1) t_{N+1-i} } + \frac{1}{N t_{1}} \Big) \hspace{-0.5mm}
	    = \hspace{-0.5mm} L,\nonumber
\end{align}
\end{small}which leads to a contradiction, where $(a)$ is due to \eqref{equ:constraint_r_1}, $(b)$ is due the expansion of $\sum_{i=0}^{N-1}x_i$ in terms of $t_{N+1-i} \sum_{j=0}^{i-1}(j+1)x_j,\ i\in[N]$, $(c)$ is due to the assumption, and $(d)$ is due to the definition of $m^{\rm (t)}$ and $\beta_i, i\in [N]$.
	Thus, we can show $\hat{\tau}(\mathbf{x},\mathbf{t}) \ge \frac{M}{N}b m^{\rm (t)}$.
	Next, it is obvious that $\mathbf{x}^{\rm (t)}$ is a feasible point and  achieves the minimum, i.e., $\hat{\tau}(\mathbf{x}^{\rm (t)},\mathbf{t}) = m^{\rm (t)}$.}
	By $\hat{\tau}(\mathbf{x},\mathbf{t}) \ge m^{\rm (t)}$ and $\hat{\tau}(\mathbf{x}^{\rm (t)},\mathbf{t}) = m^{\rm (t)}$, we know that $\mathbf{x}^{\rm (t)}$ is an optimal solution of Problem~\ref{prob:approx_solution_time}.
\end{IEEEproof}

$\mathbf{x}^{\rm (t)}$ can be interpreted as an optimal solution for a distributed computation system where $N$ workers have deterministic CPU cycle times $\mathbf{t} = \left(\mathbb{E}[T_{(n)}]\right)_{n\in[N]}$. 
Given $\mathbf{t}$, the computational complexity for calculating $\mathbf{x}^{\rm {(t)}}$ is $\mathcal{O}(N)$.



Then, we approximate the objective function of Problem~\ref{prob:relaxed_prob} by replacing random vector $\mathbf{T}$ with deterministic vector $\mathbf{t}'\triangleq (t'_n)_{n\in[N]}$, \wqm{where} $t'_n \triangleq 1 \Big/ \mathbb{E}\left[\frac{1}{T_{(n)}}\right],n\in[N]$ (which can be numerically\hspace{-0.02cm} computed\hspace{-0.02cm} for\hspace{-0.02cm} a \hspace{-0.02cm}general\hspace{-0.02cm} distribution of $T_n,n\hspace{-0.08cm}\in \hspace{-0.08cm} [N]$).

\begin{Prob}[Approximation of Problem~\ref{prob:relaxed_prob} at $\mathbf{t}'$]\label{prob:approx_solution_speed}
	\begin{align}
		\mathbf{x}^{\rm (f)} \triangleq \argmin_{\mathbf{x}} \ \  &\hat{\tau}(\mathbf{x},\mathbf{t}') \nonumber\\
		  \rm{s.t.}\ \  &\eqref{equ:constraint_r_1},\eqref{equ:constraint_r_3}.\nonumber
	\end{align}
\end{Prob}

\begin{Thm} [Closed-form Optimal Solution of Problem~\ref{prob:approx_solution_speed}]\label{thm:approx_solution_speed}
\begin{small} 
	\begin{align*}
		&x^{\rm (f)}_0 \hspace{-0.05cm} = \hspace{-0.05cm} \frac{1} {t_N'}m^{\rm (f)} \hspace{-0.1cm},\ x^{\rm (f)}_n \hspace{-0.05cm} = \hspace{-0.05cm} \frac{1}{n+1}\hspace{-0cm}\Bigg(\hspace{-0cm}\frac{1} {t_{N-n}'} \hspace{-0.05cm} - \hspace{-0cm} \frac{1} {t_{N+1-n}'}\hspace{-0cm}\Bigg)\hspace{-0.05cm}m^{\rm (f)} \hspace{-0.1cm},\ n\in[N-1],
	\end{align*}
\end{small}where $m^{\rm (f)} \triangleq \frac{L} {\sum_{n=1}^{N-1} \frac{1}{n(n+1)t_{N+1-n}'} + \frac{1}{N t_{1}'}}$.
\end{Thm}
\begin{IEEEproof}
The proof is similar to that of Theorem~\ref{thm:approx_solution_time} and is omitted due to page limitation.	
\end{IEEEproof}

Let $F_n \triangleq \frac{1}{T_n}, n\in[N]$ denote the CPU frequencies of the $N$ workers.
Thus, $\mathbf{x}^{\rm (f)}$ can be interpreted as an optimal solution for a distributed computation system where $N$ workers have deterministic CPU frequencies $\left(\mathbb{E}[F_{(n)}]\right)_{n\in[N]}$. 
Given $\mathbf{t}'$, the computational complexity of calculating $\mathbf{x}^{\rm {(f)}}$ is $\mathcal{O}(N)$.


\subsection{Analysis of Approximate Solutions}
\label{ssec:analysis_of_approx_solution}

In this part, we characterize the two approximate solutions under the assumption that $T_n,n\in[N]$ are i.i.d. according to a shifted-exponential distribution, i.e., $\Pr[T_n\le t]=1-e^{-\mu(t-t_0)}, t\ge t_0$, where $\mu$ is the rate parameter and $t_0$ is the shift parameter. 
Notice that shifted-exponential distributions are widely considered in modeling stragglers in a distributed computation system~\cite{Lee_speeding_up,Draper_hierarchical_ISIT,Draper_hierarchical_CSIT,Min_Ye_communication   }.

First, we derive the expressions of $\mathbf{t}$ and $\mathbf{t}'$ which are parameters of the two approximate solutions, respectively.
By~\cite{renyi1953theory}, we have: 
\begin{align}
	t_n = \frac{1}{\mu}(H_N-H_{N-n})+t_0,n\in[N],\label{equ:Renyi}
\end{align}
where $H_n\triangleq\sum_{i=1}^n \frac{1}{n}$ is the $n$-th harmonic number.
\begin{Lem}
\label{lem:t'}
If $t_0>0$,\footnote{When $t_0=0$, $E_i(0)$ does not exist.} for all $n\in[N]$, we have $t'_n$ given in \eqref{equ:t'} as shown at the top of this page,
where $E_i(x)\triangleq \int_{-\infty}^x \frac{e^t}{t}dt$ is the exponential integral.
\end{Lem}
\begin{IEEEproof}
	By the probability density function of order statistics, we have $\frac{1}{t'_{n}}= -\mu n\binom{N}{n-1}\int_0^1 \frac{x^{N-n}(1-x)^{n-1}}{\log(x)-\mu t_0} dx$.
	\Globecomb{Letting} $I_{t_0}(p,q) \triangleq \int_0^1 \frac{x^{p-1} (1-x)^{q-1}}{\log(x)-\mu t_0} dx$, \Globecomb{where $p,q$ are positive integers,
	we have $\frac{1}{t'_{n}}= -\mu n\binom{N}{n-1} I_{t_0}(N-n+1,n)$.
	By noting that $I_{t_0}(p,q) = I_{t_0}(p+1,q) + I_{t_0}(p,q+1)$, we can show that $I_{t_0}(p,q) = \sum_{i=0}^{1-1}(-1)^i \binom{1-1}{i}e^{\mu t_0 (p+i)} E_i(-\mu t_0 (p+i))$, by induction on $q$ for fixed $p$.
	Therefore, we complete the proof}.
\end{IEEEproof}

Note that the computational complexities for calculating the parameters for $\mathbf{x}^{\rm (t)}$ and  $\mathbf{x}^{\rm (f)}$, i.e., $\mathbf{t}$ and $\mathbf{t}'$, are $\mathcal{O}(N)$ and $\mathcal{O}(N^2)$, respectively.
Then, we characterize the sub-optimalities of $\mathbf{x}^{\rm (t)}$ and  $\mathbf{x}^{\rm (f)}$, respectively.
Recall that $\hat{\tau}^*_{\rm avg-ct}$ denotes the optimal value of Problem~\ref{prob:relaxed_prob}.
\begin{Thm}[Sub-optimality Analysis]
\label{thm:multiplicative_gap}
\begin{small}
\begin{align*}
	\frac{\mathbb{E}_{\mathbf{T}}\left[\hat{\tau}(\mathbf{x}^{\rm (t)},\mathbf{T})\right]} {\hat{\tau}^*_{\rm avg-ct}} = \mathcal{O}\left(\left(\Acceptb{\log N}\right)^2\right),
	\frac{\mathbb{E}_{\mathbf{T}}\left[\hat{\tau}(\mathbf{x}^{\rm (f)},\mathbf{T})\right]} {\hat{\tau}^*_{\rm avg-ct}} = \mathcal{O}\left(\Acceptb{\log N}\right).
\end{align*}
\end{small}
\end{Thm}
\begin{IEEEproof}
	First, by Jensen's inequality and the definition of $\mathbf{x}^{\rm (t)}$, we have $\hat{\tau}^*_{\rm avg-ct} \ge \hat{\tau}(\mathbf{x}^{\dag},\mathbf{t}) \ge \hat{\tau}(\mathbf{x}^{\rm (t)},\mathbf{t})$.
	Then, by $\hat{\tau}^*_{\rm avg-ct} \ge \hat{\tau}(\mathbf{x}^{\rm (t)},\mathbf{t})$ and \eqref{equ:Renyi}, we have $\frac{\mathbb{E}_{\mathbf{T}}\left[\hat{\tau}(\mathbf{x}^{\rm (t)},\mathbf{T})\right]} {\hat{\tau}^*_{\rm avg-ct}} \le \frac{(H_N+1)(H_N+\mu t_0)}{\mu^2 t_0^2} = \mathcal{O}\left(\left(\Acceptb{\log N}\right)^2\right)$.
	Finally, by $\hat{\tau}^*_{\rm avg-ct} \ge \hat{\tau}(\mathbf{x}^{\rm (t)},\mathbf{t})$,  Lemma~\ref{lem:t'} and Cauchy–Schwarz inequality, we have $\frac{\mathbb{E}_{\mathbf{T}}\left[\hat{\tau}(\mathbf{x}^{\rm (f)},\mathbf{T})\right]} {\hat{\tau}^*_{\rm avg-ct}} \le \frac{H_N}{\mu t_0}+1 = \mathcal{O}\left(\Acceptb{\log N}\right)$. 
\end{IEEEproof}

Theorem~\ref{thm:multiplicative_gap} indicates that for any $L$, the expected overall runtimes of the two approximate solutions $\mathbf{x}^{\rm (t)}$ and $\mathbf{x}^{\rm (f)}$ and the minimum expected overall runtime have sub-linear multiplicative gaps in $N$.
\wqr{As $\left(\Acceptb{\log N}\right)^2=4$ requires $N=8$ and $\Acceptb{\log N}=4$ requires $N=55$, the analytical upper bounds on the gaps are not large at a small or moderate $N$}. 
Later in Sec.~\ref{sec:Numerical_Results}, we shall see that the actual gaps are very small even at $N=50$.
Notice that the multiplicative gap for $\mathbf{x}^{\rm (f)}$ is smaller, but the computational complexity for calculating parameters $\mathbf{t}'$ for $\mathbf{x}^{\rm (f)}$ is higher.

\section{Numerical Results}
\label{sec:Numerical_Results}

\Acceptb{In this section}, we compare the integer-valued approximations of the proposed solutions in Sec.~\ref{sssec:solution} and Sec.~\ref{ssec:approx_solution}, \wqm{denoted by $\hat{\mathbf{x}}^{\dag}$, $\hat{\mathbf{x}}^{\rm (t)}$ and $\hat{\mathbf{x}}^{\rm (f)}$ (obtained using the rounding method in~\cite[pp. 386]{Boyd_cvxbook})} with four baseline schemes derived from  \cite{Tandon_gradientcoding,Draper_hierarchical_ISIT}.
Specifically, single-block coordinate gradient coding (BCGC)  corresponds to the integer-valued solution obtained by solving Problem~\ref{prob:equivalent_1} with extra constraints $\|\mathbf{x}\|_0=1$ and rounding the optimal solution using the rounding method in~\cite[pp. 386]{Boyd_cvxbook}.
Notice that single-BCGC can be viewed as an optimized version of the gradient coding scheme for full stragglers in~\cite{Tandon_gradientcoding}.
Tandon \emph{et al}.'s gradient coding corresponds to the optimal gradient coding scheme in~\cite{Tandon_gradientcoding} for $\alpha$-partial stragglers with $\alpha=\frac{ \mathbb{E}[T_n|T_n\le t] } {  \mathbb{E}[T_n|T_n> t] }=6$, where $t$ satisfies $\Pr[T_n\le t]=0.5$.
Ferdinand \emph{et al}'s coded computation ($r=L$) and ($r=L/2$) correspond to the optimal coding scheme in~\cite{Draper_hierarchical_ISIT}, with the optimized coding parameter at the number of layers $r=L$ and $r=L/2$, respectively.
\Acceptb{In the simulation, $T_n, n\in[N]$ follow the shifted-exponential distribution with $\mu$ and $t_0=50$.
We set $M=50$ and $b=1$.}

\begin{figure}[t]
\begin{center}
	{\resizebox{4.8cm}{!}{\includegraphics{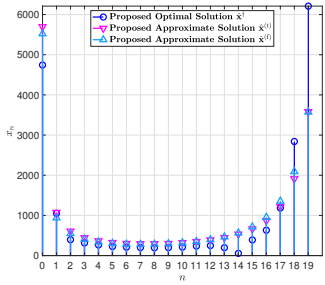}}}
\end{center}
  \vspace{-0.1cm}
  \caption{$\hat{\mathbf{x}}^{\dag}$, $\hat{\mathbf{x}}^{\rm (t)}$ and $\hat{\mathbf{x}}^{\rm (f)}$ at $N=20$, $L=2\times 10^4$ and $\mu=10^{-3}$.}\label{fig:x_vs_n}
  \vspace{-0.1cm}
\end{figure}

\begin{figure}[t]
\begin{center}
  \subfigure[\scriptsize{$L=2\times 10^4$ and $\mu=10^{-3}$.}\label{subfig:tau_vs_N}]
  {\resizebox{4.3cm}{!}{\includegraphics{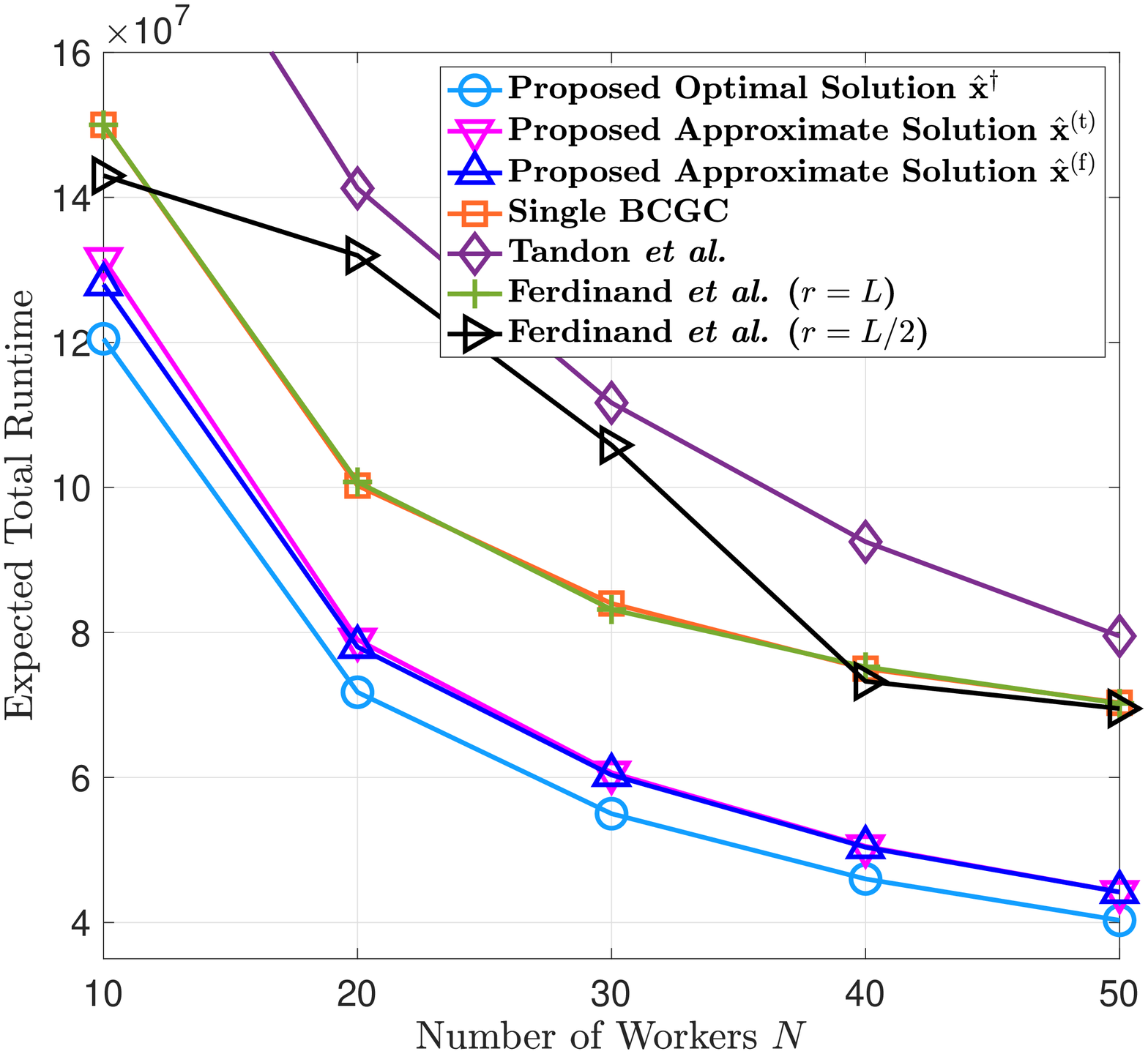}}}
  \subfigure[\scriptsize{$N=20$ and $L=2\times 10^4$.}\label{subfig:tau_vs_mu}]
  {\resizebox{4.3cm}{!}{\includegraphics{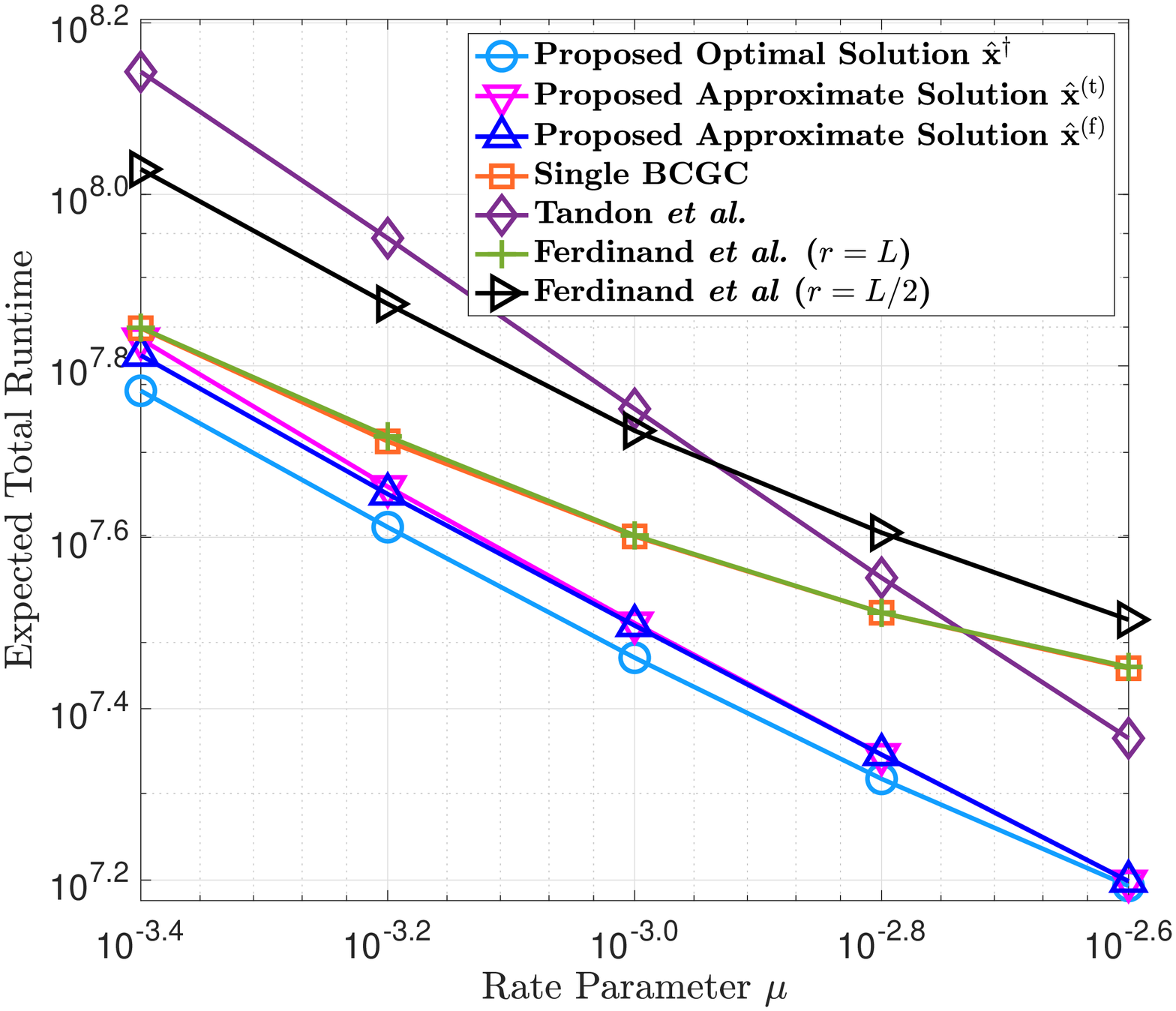}}}
  \end{center}
         \caption{Expected total runtime versus $N$ and $\mu$. 
         }\label{fig:simulation}
\end{figure}

\nwqc{Fig.~\ref{fig:x_vs_n} illustrates the proposed solutions $\hat{\mathbf{x}}^{\dag}$, $\hat{\mathbf{x}}^{\rm (t)}$ and $\hat{\mathbf{x}}^{\rm (f)}$, respectively.
Fig.~\ref{fig:x_vs_n} indicates that in these solutions, the first block (containing coordinates $1,\cdots,x_0$) with no redundancy and the last block (containing coordinates $L-x_{N-1}+1,\cdots,L$) with redundancy for tolerating $N-1$ stragglers contain most of the $L$ coordinates.}
Fig.~\ref{subfig:tau_vs_N} and Fig.~\ref{subfig:tau_vs_mu}  illustrate the expected runtime versus the number of workers $N$ and the rate parameter $\mu$, respectively.
From Fig.~\ref{subfig:tau_vs_N}, we see that the expected overall runtime  of each scheme decreases with $N$, due to the increase of the overall computation resource with $N$.
From Fig.~\ref{subfig:tau_vs_mu}, we see that the expected overall runtime of each scheme decreases with $\mu$ due to the decrease of $\mathbb{E}[T_n]=\frac{1}{\mu}+t_0$ with $\mu$.
Furthermore, from Fig.~\ref{fig:simulation}, we can draw the following conclusions.
The proposed solutions significantly outperform the four baseline schemes.
For instance, the proposed solutions can achieve reductions of 37\% and 44\% in the expected overall runtime over the best baseline scheme at $N=50$ in Fig.~\ref{subfig:tau_vs_N} and at $\mu=10^{-2.6}$ in Fig.~\ref{subfig:tau_vs_mu}, respectively.
The gains over single BCGC and Tandon \emph{et al}.'s gradient coding are due to the diverse redundancies introduced across \wqm{partial derivatives}.
The gains over Ferdinand \emph{et al}'s coded computation schemes at $r=L$ and $r=L/2$ indicate that an optimal coded computation scheme for calculating matrix-vector multiplication is no longer effective for calculating a general gradient.
The proposed closed-form approximate solutions are quite close to the proposed optimal solution and hence have significant practical values.
Besides, note that \Acceptb{$\hat{\mathbf{x}}^{\rm (f)}$} slightly outperforms \Acceptb{$\hat{\mathbf{x}}^{\rm (t)}$} in accordance with Theorem~\ref{thm:multiplicative_gap}.

\section{Conclusion}
\label{sec:Conclusion}

In this paper, we propose an optimal block coordinate gradient coding scheme, providing $N$ levels of redundancies for tolerating $0, 1,\cdots, N-1$ stragglers, respectively, based on a stochastic optimization problem to minimize the expected overall runtime for collaboratively computing the gradient. We obtain an optimal solution using a stochastic projected subgradient method and propose two low-complexity approximate solutions with closed-from expressions, for the stochastic optimization problem. We also show that under a shifted-exponential distribution, \wqm{for any $L$}, the expected overall runtimes of the two approximate solutions and the minimum overall runtime have sub-linear multiplicative gaps \wqm{in $N$}. Finally, numerical results show that the proposed solutions significantly outperform existing ones, and the proposed approximate solutions achieve close-to-minimum expected overall runtimes.


\bibliographystyle{IEEEtran}
\bibliography{refs_wq.bib}

\end{document}